\documentclass[letter, twocolumn]{jpsj3}
\usepackage{txfonts}
\usepackage{braket}
\usepackage{bm}

\usepackage{graphicx}
\usepackage{dcolumn}
\usepackage{bm}
\usepackage{amsmath}
\usepackage{times}
\usepackage[usenames]{color}
\title{Photo-Induced Phase Transition in Charge Order Systems \\ --Charge Frustration and Interplay with Lattice--}

\author{Hiroshi Hashimoto$^1$\thanks{hashimoto@cmpt.phys.tohoku.ac.jp}, Hiroaki Matsueda$^2$, Hitoshi Seo$^{3,4}$, and Sumio Ishihara$^{1}$}
\inst{$^1$Department of Physics, Tohoku University, Sendai 980-8578, Japan \\
$^2$Sendai National College of Technology, Sendai 989-3128, Japan \\
$^3$Condensed Matter Theory Laboratory, RIKEN, Wako, Saitama 351-0198, Japan \\
$^4$RIKEN Center for Emergent Matter Science (CEMS), Wako, Saitama 351-0198, Japan \\
}

\abst{
Lattice effects on photo-excited states in interacting charge frustrated system are examined. 
Real time dynamics in the interacting spinless fermion model on a triangular lattice coupled to lattice vibration are analyzed by applying the exact diagonalization method combined with the classical equation of motion. 
A photo-induced phase transition from the horizontal stripe-type charge order (CO) to the 3-fold CO occurs through a characteristic intermediate time domain. 
By analyzing the time evolution in detail, we find that this characteristic dynamics are seen when the electron and lattice sectors are not complementary to each other but show cooperative time evolutions. 
We also find that there are threshold values in the optical fluorescence and electron-lattice coupling for emergence of the photo-induced 3-fold CO.  
The dynamics are distinct from those from the vertical stripe-type CO, in which a monotonic CO melting occurs. 
A scenario of the photo-induced CO phase transition with lattice degree of freedom is presented from a view point of charge frustration. 
}

\begin{document}
\maketitle

Emergence of inequivalent electronic charge density and its long range order in a solid are termed electronic charge order (CO). 
It is observed in a wide class of solids, e.g. transition-metal oxides and organic molecular materials. 
In these systems, a rich variety of exotic phenomena are found on the verge of CO,
such as metal-insulator transition, colossal magnetoresistance, superconductivity and so on. ~\cite{yamamoto, tomioka, axe} 
Regarding the real-space CO pattern, lattice geometry plays an essential role. 
In particular, geometrical frustration effects on the interacting charge systems have been long-standing issues.~\cite{anderson, verwey} 
In such charge frustrated systems, phase competition and the existence of a macroscopic number of degenerated charge configurations are 
expected to affect not only the stabilization of CO but also the related phenomena realized in it. 

In the meantime, there have been recent developments in artificially controlling the CO states by external field, using different experimental tools. 
Among such external stimuli, ultrafast optical pulse has been recognized as a powerful tool of 
the CO control technique. 
Not only the optically induced CO melting, but also emergence of a hidden CO state and other exotic optically induced phenomena have been observed experimentally. ~\cite{kawakami, ishikawa, onda, gao, beaud} 
The CO systems are recognized as suitable targets in such ultrafast optical studies, since  
the CO insulators are more susceptible than insulators owing to other mechanisms such as the Mott insulators. 
In the transient optical responses in CO materials, 
it is widely accepted that lattice degrees of freedom play a key ingredient. 
First of all, the characteristic time scales for the electron and lattice degrees of freedom are different, the former being faster than the latter in general.  
Then, the electron-lattice coupling gives an energy flow from the electron sector to the lattice sector.
Furthermore, by separating the electron-electron and electron-lattice interactions, having different characteristic time scales, 
transient optical study may identify the dominant stabilization mechanism of the CO at the equilibrium.
Recently developed time-resolved diffraction and spectroscopy techniques greatly promote 
direct observations of the transient lattice dynamics in the electron-lattice coupled CO state.
In order to study such issues, theoretical calculations for the photo-induced CO systems in which the lattice degrees of freedom are taken into account have been performed so far. ~\cite{yonemitsu, kawakami, onda, dagotto, yonemitsu2} 
Combination effects on the electron-lattice interactions and geometrical frustration should be important, since they both underlie many CO materials especially molecular compounds~\cite{merino}.
Nevertheless, these effects on the photo-excited transient processes and its mechanism are not fully clarified yet. 

\begin{figure}[t]
\includegraphics[width=\columnwidth,clip]{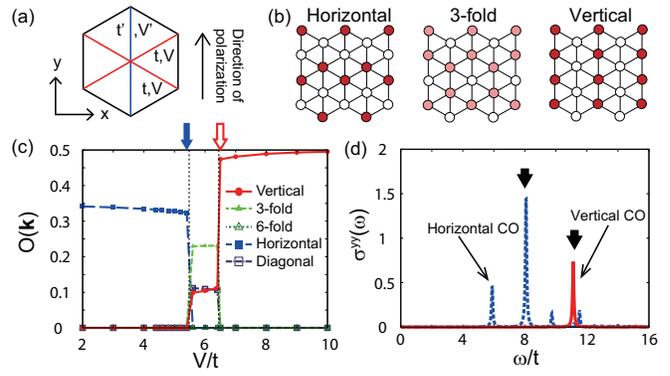}
\caption{
(Color online) 
(a) Fermion hoppings and Coulomb interactions on a triangular lattice. 
(b) Schematic charge configurations for the h-stripe, 3-fold and v-stripe CO structures.
Filled and open circles represent charge rich and poor sites, respectively.
(c) The $V$-dependence of the charge order parameters in the ground state with keeping $V+V'=12t$.
Filled and open arrows indicate parameter values chosen in Figs.~\ref{fig:tran1}(a) and (b), respectively.
(d) The optical conductivity spectra before photo-excitation in the h-stripe and v-stripe CO phases.
Bold arrows represent the energies which are chosen as energies of the pump photons.
Parameter values for the lattice are $\omega_{lat}/t=0.25$ and $\lambda/t=0.8$.
}
\label{fig:gs}
\end{figure}

In this Letter, we examine lattice effects on the photo-exited CO states in a geometrically  frustrated lattice. 
We analyze the interacting spinless fermion model on a triangular lattice where local lattice vibrations and the Holstein-type electron-lattice interaction are taken into account.  
This model without the lattice degree of freedom show three kinds of COs in the ground state, according to the anisotropy of the Coulomb interactions,
i.e. the horizontal stripe-type (h-stripe), vertical stripe-type (v-stripe) and 3-fold CO states [see Fig.~\ref{fig:gs}(b)]. ~\cite{hotta, nishimoto0, nishimoto, naka} 
Photo-induced electron dynamics in this model were examined in our  previous paper.~\cite{hashimoto} 
It was shown that the photo-induced phase change occurs from the h-stripe to 3-fold COs, which is attributable to the charge frustration effects in the photo-excited states. 
Here, by adding the electron-lattice interaction, we will see in the following that the photo-induced transition from the h-stripe to 3-fold COs occurs through an intermediate time domain,
where the electron and lattice sectors are not complementary to each other, but show cooperative time evolutions. This is in contrast to the photo-excitation in the v-stripe CO, as well as that in the h-stripe CO without the lattice degree of freedom. 
Based on the numerical results, we present a scenario for the photo-induced phase transitions in charge frustrated electron-lattice coupled systems.

The Hamiltonian is given by 
\begin{align}
 {\cal H}=& -\sum_{\braket{ij}}t_{ij} c_{i}^{\dagger}c_{j} 
+ \sum_{\braket{ij}} V_{ij} n_{i}n_{j}
\nonumber \\
&+ \frac{\omega_{lat}}{2}\sum_{i}( q_{i}^{2}+p_{i}^{2}) -\lambda \sum_{i}q_{i} n_{i} , 
\label{eq:e1}
\end{align}
where $c_{i}$ ($c_{i}^{\dagger}$) is an annihilation (creation) operator for a spinless fermion at site $i$, 
and $n_{i}=c_{i}^{\dagger}c_{i}$ is a number operator.
We introduce local lattice vibrations where the displacement and its conjugated momentum at site $i$ are written as $q_{i}$ and $p_{i}$, respectively. 
The first and second terms represent the fermion hoppings and the inter-site 
Coulomb interactions, respectively. 
As shown in Fig.~\ref{fig:gs}(a), 
anisotropies in the fermion hoppings and the Coulomb interactions are introduced as $(t, t')$, and $(V, V')$, respectively. 
The lattice vibration with frequency $\omega_{lat}$ is represented in the third term, and the Holstein-type electron-lattice coupling is given in the last term where $\lambda$ is a coupling constant. 

As in our previous work~\cite{hashimoto}, the optical pump pulse is introduced as the Peierls phase into the transfer integral as 
$t_{ij}\rightarrow t_{ij}e^{-{\rm i}\bm{A}(\tau)\cdot \bm{R}_{ij}}$
where ${\bm A}(\tau)$ is the vector potential at time $\tau$,
and $\bm{R}_{ij}$ is a relative position vector connecting sites $i$ and $j$, which we take a Gaussian form given by 
$\bm{A}(\tau) = A_{p}\bm{e}(\sqrt{2\pi}\tau_{p})^{-1}
\exp[-\tau^{2}/(2\tau_{p}^{2})] \cos(\omega_{p}\tau)$ 
with amplitude $A_{p}$, frequency $\omega_{p}$, a damping factor $\tau_p$, and a unit vector $\bm{e}$. 

Steady and transient properties are calculated in finite size clusters where the cluster sizes are taken up to $N=4\times 6=24$ sites and the periodic boundary condition is imposed. 
The electronic wave functions $\ket{\Psi(\tau)}$ are calculated by the Lanczos-based exact diagonalization method ~\cite{park, prelovsek} as in our previous work~\cite{hashimoto} with a time step $\delta \tau$. 
%
As for the lattice part, the Newtonian equation given by
$\ddot{q}_{i}=-\omega_{lat}^{2}q_{i}+\omega_{lat}\lambda\braket{n_{i}}$ 
is solved numerically by using the leap-frog method, 
$q_{i}(\tau+\delta\tau') = q_{i}(\tau) + \delta \tau' p_{i}(\tau+\delta\tau'/2) $, and 
$ p_{i}(\tau+\delta\tau'/2)=p_{i}(\tau-\delta\tau'/2) + \delta \tau' F_i(\tau) $. 
We introduce a time step $\delta \tau '$, and the Hellmann-Feynman force $F_i(\tau)= -\omega_{lat}q_{i}(\tau) + \lambda \braket{n_{i}}_\tau$ where $\braket{\cdots}_\tau$ is an expectation value calculated by $\ket{\Psi(\tau)} $. 
Validity of the present treatment for the lattice degree of freedom is checked by comparing the results with the ones calculated by the method proposed in Ref.~\citen{gomi2}.
%
In the numerical calculations, the number of the fermions are fixed to be $N/2$, 
the parameters are chosen as $t=t'$, $V+V'=12t$, and $\tau_p = 3/t$, and 
the polarization of the pump photon is set parallel to the $y$ direction.
We choose $M=15$, $\delta \tau=0.01 /t$ and $\delta\tau'= \delta \tau \times 10^{-4}$ which are sufficient to obtain the results with high enough accuracy~\cite{initial}. 

Let us start from the physical properties before photo-excitation.
In Fig.~\ref{fig:gs}(c), we plot the charge order parameters defined by
$O(\bm{k}) = N^{-1}|\sum_{i} \braket{(n_{i}-1/2)}e^{{\rm i}\bm{k}\cdot\bm{R}_{i}}|$
as a function of $V$ for $\omega_{lat}/t=0.25$ and $\lambda /t=0.8$.
Finite values of $O(\bm{k})$ are owing to spatial symmetry breaking due to the lattice distortions.
The result is very similar to the previous results without the lattice degree of freedom.
In the limiting regions, two prototypical stripe COs are seen, 
i.e. the h-stripe and v-stripe COs
characterized by the wave vectors, $\bm{k}_{\rm H}\equiv (0, \pi) $, and $ {\bm k}_{\rm V} \equiv (2\pi/\sqrt{3}, 0) $, respectively (see Fig.~\ref{fig:gs}(b)).
In between the two COs around the frustration points, $V/t \sim V'/t \sim 6$, 
the 3-fold CO characterized by ${\bm k}_{\rm 3}\equiv (2\pi/\sqrt{3}, 2\pi/3)$ 
appears.  
The phase boundary points are almost the same for the cases with or without the lattice, while the amplitudes of the diagonal and vertical COs are slightly larger here. 
The lattice distortions follow the charge distribution patterns since in the steady state they are just proportional to each other. 

The optical spectra before pumping are shown in Fig.~\ref{fig:gs}(d) where the h-stripe and v-stripe COs close to the phase boundaries are chosen ($V/t=5.4$, and $6.5$) . 
The regular part of the optical conductivity is given by 
$
\sigma^{\alpha \alpha}(\omega) = -(N\omega)^{-1} 
\text{Im}
\bra{\Psi_{0}} j^{\alpha} 
(\omega-{\cal{H}}+E_{0}+{\rm i}\eta)^{-1}
j^{\alpha} 
\ket{\Psi_{0}} 
$
where $\ket{\Psi_0}$ and $E_0$ are the electronic wave function and energy in the ground-state, respectively, $j^{\alpha}$ is a current operator with a Cartesian coordinate $\alpha(= x, y)$, and $\eta$ is an infinitesimal constant. 
We adopt the adiabatic approximation in the calculation of the optical spectra, where the lattice coordinates are fixed. 
As the pump photon energies in the time evolutions, 
we chose the energies at which the spectra take their maxima shown by arrows in Fig.~\ref{fig:gs}(d). 

\begin{figure}[t]
\begin{center}
\includegraphics[width=0.7\columnwidth,clip]{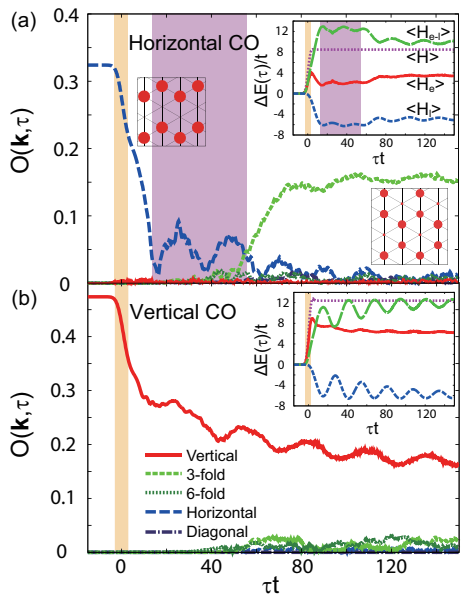}
\end{center}
\caption{
(Color online)
(a) Time dependences of the charge order parameters in the h-stripe and (b) v-stripe cases.
Real-space charge distributions before pumping ($\tau t=-15$) and at $\tau t =120$ are also shown in (a).
Insets show time dependences of several components of the energy.
Orange light and purple dark shaded areas represent the time intervals when the pump pulses are introduced
and the intermediate time domain, respectively.
Parameter values are chosen to be $V/t=5.4$, $\omega_p/t=8$ and $A_p=0.8$ in (a) and (b),
and $V/t=6.5$, $\omega_p/t=11.2$ and $A_p=1$ in (c) and (d).
Other parameter values are $\omega_{lat}/t=0.25$ and $\lambda /t=0.8$.
}
\label{fig:tran1}
\end{figure}
%
%
From here, we consider the two cases, the h-stripe case (the photo-excitation to the h-stripe CO) and the v-stripe case (that to the v-stripe CO).
The time dependences of the charge order parameters in the h-stripe and v-stripe cases are shown in Fig.~\ref{fig:tran1}(a) and (b), respectively. 
In both cases, the initial COs are weakened after the photo-irradiation. 
As shown in Fig.~\ref{fig:tran1}(a), in the h-stripe case, not only its melting but also emergence of the 3-fold CO are observed, where finally the dominant CO parameter is interchanged. 
Characteristically, one can see that there are three time domains; the initial CO melts monotonically in $\tau<15/t$, 
amplitudes of all COs are small in $15/t < \tau< 60/t$, and finally, 
$O({\bm k}_3)$ develops and tends to be saturated in $60/t < \tau$. 
This characteristic time evolution will be discussed later in more detail. 
In clear contrast, in the v-stripe case, shown in Fig.~\ref{fig:tran1}(b),
$O({\bm k}_V)$ just monotonically decreases after the photo-irradiation.
These transient behaviors are not observed in our previous work for the purely electronic model~\cite{hashimoto},
and therefore are attributed to the energy relaxation to the lattice system. 

Energy flows in the electron and lattice degrees of freedom give a hint to understand 
contrasting behaviors above. 
Several components of energies in the two cases are shown in the insets of Figs.~\ref{fig:tran1}(a) and (b).
We define $\braket{{\cal H}_{\rm e}}$, $\braket{{\cal H}_{\rm l}}$, and  $\braket{{\cal H}_{\rm e-l}}$, respectively,
as the expectation values of the sum of the first and second terms, the third term, and the fourth term of the Hamiltonian in Eq.~(\ref{eq:e1}). 
During the time interval when the pump pulse is applied, i.e. $\tau < \tau_{d} $, 
increases in $\braket{{\cal H}_{\rm e}}$ and  $\braket{{\cal H}_{\rm e-l}}$ indicate photo-excitations of the electron sector. 
Note that the conservation of the total energy in $\tau > \tau_{d}$ guarantees good accuracy in the numerical calculations. 
In the v-stripe case, monotonic changes in all components of energy after photo-excitations are consistent with the results of the charge order parameters shown in Fig.~\ref{fig:tran1}(b).  
On the other hand, in the h-stripe case, at the intermediate time domain $ 15/t < \tau < 60/t$, $\braket{{\cal H}_{\rm e-l}}$ shows a plateau, and $\braket{{\cal H}_{\rm e}}$ and $\braket{{\cal H}_{\rm l}}$ show some basin-like shapes. 
This plateau behavior in $\braket{{\cal H}_{\rm e-l}}$ implies desynchronization of the electron and lattice sectors. 
Then at $\tau = 60/t$, the decrease in $\braket{{\cal H}_{\rm e-l}}$ is seen, implying a reduction of the discrepancy between 
the electron and lattice sectors owing to the compatible lattice distortion associated with the emergence of the 3-fold CO. 

\begin{figure}[t]
\includegraphics[width=\columnwidth,clip]{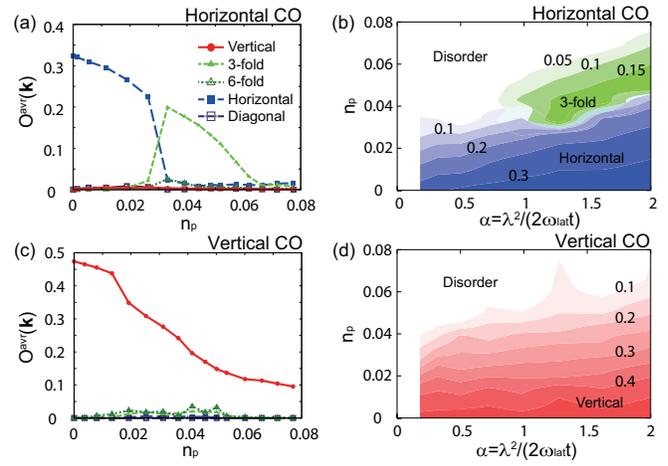}
\caption{
(Color online) 
(a) Fluorescence dependences of the charge order parameters after the photo-excitation in the h-stripe case, and (c) those in the v-stripe case as functions of the absorbed photon density.
(b) A contour map of the charge order parameters averaged in $120 \le \tau t \le 150$ in the h-stripe case, and (d) that in the v-stripe case. 
Pump photon amplitudes are adopted in the range of $0 \le  A_p \le 2$.
Other parameter values in (a) and (c) are the same as those in Figs.~\ref{fig:tran1}(a) and (b), respectively.
}
\label{fig:map}
\end{figure}
Next, we investigate the fluorescence dependences, which clearly demonstrate the lattice effects on the photo-induced transition from the h-stripe CO to the 3-fold CO. 
In Figs.~\ref{fig:map}(a) and (c),  respectively, we show the averaged order 
parameters ($O^{\text{av}}(\bm{k})$) in the h- and v-stripe cases as functions of the absorbed photon density. 
We define $O^{\text{av}}(\bm{k})$ as the order parameter averaged in the time domain of $120 \le \tau t \le 150$. 
We measure the absorbed photon density by introducing $n_{p} = (E^{\text{av}}-E_{0})/(N\omega_{p})$, where $E^{\text{av}}$ is the total energy averaged in $120 \le \tau t \le 150$,
and $E_{0}$ is the total energy before pumping.
In the v-stripe case (Fig.~\ref{fig:map}(c)),
the order parameter decreases monotonically with increasing $n_{p}$ as expected. 
On the other hand, in the h-stripe case, a first order transition like behavior occurs from the h-stripe CO to the 3-fold CO 
around $n_{p} \sim 0.03$, implying the existence of the threshold fluorescence. 
%
Phase diagrams in the photo-excited states are presented as contour maps of $O^{\text{av}}(\bm{k})$
in the $n_p$-$\alpha (\equiv \lambda^{2}/(2\omega_{lat}t) )$ planes in Figs.~\ref{fig:map}(b) and (d). 
The melting of  the v-stripe CO is suppressed in regions of small $n_p$, whereas it is rather insensitive to the value of $\alpha$; the melting is mainly of an electronic origin. 
As seen in Fig.~\ref{fig:map}(b), the 3-fold CO phase appears in between the h-stripe CO and a disordered state in large $\alpha$.  
There is a threshold electron-lattice coupling for the photo-induced 3-fold CO. 
This implies that the electron-lattice coupling stabilizes both the h-stripe and 3-fold COs, although the lattice contribution is larger in the 3-fold CO. 
%

\begin{figure}[t]
\begin{center}
\includegraphics[width=\columnwidth,clip]{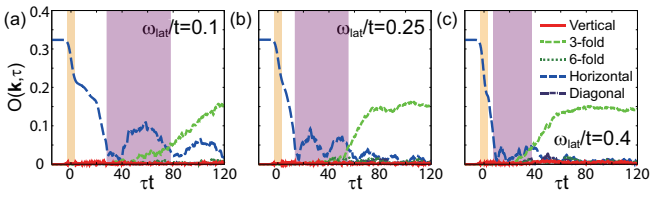}
\end{center}
\caption{
(Color online) 
Time dependences of the charge order parameters in the h-stripe case for several lattice vibration frequencies. 
Parameter values are chosen to be $\omega_{lat}/t=0.1$, $\lambda/t=0.5$, $A_p=0.8$ in (a),
$\omega_{lat}/t=0.25$, $\lambda/t=0.8$, $A_p=0.8$ in (b), and $\omega_{lat}/t=0.4$, $\lambda/t=1$, $A_p=1.2$ in (c).
Other parameter values are  $V/t=5.4$ and $\omega_p/t=8$.
%
}
\label{fig:adia}
\end{figure}

Here we discuss the vibration frequency dependence of the time interval where the intermediate time domain appears in the photo-excitation from the h-stripe CO to the 3-fold CO.
In Fig.~\ref{fig:adia}, the time dependences of $O(\bm{k})$ for several $\omega_{lat}$ are presented.
It is clearly shown that the intermediate time domain shrinks with increasing $\omega_{lat}$. 
That is, the time interval is determined by the vibration frequency,  and the lattice vibration acts as a driving force of the h-stripe CO to the 3-fold CO. 
The decrease in $\omega_{lat}$ hardly affects the amplitude of the induced $O(\bm{k}_{3})$,
while it enhances the lattice displacement.

\begin{figure}[t]
\includegraphics[width=\columnwidth,clip]{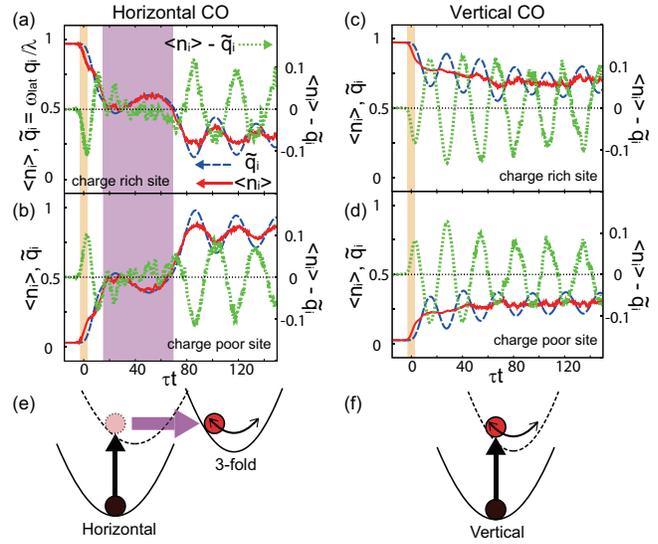}
\caption{
(Color online) 
(a) Time dependences of the charge density ($\braket{n_i}$), the normalized lattice displacement ($\tilde{q}_i$), 
and difference between the two ($\braket{n_{i}}-\tilde{q}_{i}$) at a charge rich site, and (b) those at a charge poor site in the h-stripe case.
Results in the v-stripe case are shown in (c) and (d). 
Parameter values in (a) and (b) are chosen to be $V/t=5.4$, $\omega_{lat}/t=0.25$, $\lambda/t =1$, $\omega_p/t=9.6$ and $A_p=1.2$, 
and those in (c) and (d) are the same as those in Fig.~\ref{fig:tran1}(b). 
(e) Schematic adiabatic potentials in the h-stripe case, and (f) those in the v-stripe case. 
}
\label{fig:site}
\end{figure}

Such pictures are consistent with the time dependences of the local electronic charge density, $\braket{n_i}$, and the normalized lattice displacement, $\tilde{q}_i \equiv \omega_{lat} q_{i}/\lambda$. 
%
As shown in Figs.~\ref{fig:site}(a) and (b), $\braket{n_i}$ and $\tilde{q}_i$ in the h-stripe case show almost similar time dependence which is featureless in  the intermediate time domain 
(We slightly varied the parameters from the results above to see the characteristics clearly. Now the time domain is $ 15/t < \tau < 70/t$).  
On the other hand, in $0 < \tau < 15/t$ and $\tau > 70/t$ in the h-stripe case, and in all time regions after photo-excitation in the v-stripe case (see Figs.~\ref{fig:site}(c) and (d)), 
$q_i$ shows clear oscillation, while oscillatory behavior in $\braket{n_i}$ is weak. 
%
The difference $\braket{n_{i}}-\tilde{q}_{i}$ plotted by green dotted curves in Figs.~\ref{fig:site}
show small values in the intermediate time domain in the h-stripe case, and is distinguishable from other time domains. 
Therefore, the electron and lattice sectors do change cooperatively in the intermediate time domain observed in the h-stripe case,
while, in others, the electronic states are almost settled, and the lattice oscillates coherently around a stable position, which is almost determined by the photo-excited meta-stable electronic states. 

The obtained characteristic time evolutions are interpreted qualitatively by the adiabatic potential schemes shown in Figs.~\ref{fig:site}(e) and (f). 
In the v-stripe case, the photo-excited state is a charge melted metallic-like state. 
In the adiabatic potential for the photo-excited state, lattices oscillate around the stabilization point. 
On the other hand, in the h-stripe case, the Frank-Condon photo-excited state turns into the 3-fold CO state. 
In the translocation processes between the two CO states, the electron and lattice degrees of freedom are not complementary to each other, and the electron-lattice interaction energy is high. 
This discrepancy induces a driving force of the phase transition, in which the electron and lattice sectors change cooperatively. 
After the system is settled down into the 3-fold CO, the lattice oscillates around the equilibrium position in the new adiabatic potential plane. 
Both in the early and late time domains in the transient processes, a picture for the lattice vibrations on a adiabatic potential is valid and the electronic energy scale is much higher than the lattice energy scale. 
On the other hand, in the intermediate time domain, the two energy scales are close with each other and the non-adiabatic processes occur. 
This small electronic energy scale is attributable to the frustration effect in the photo-excited state in the h-stripe CO~\cite{hashimoto}. 

We discuss a relation of the present results to our previous work without the lattice degree of freedom~\cite{hashimoto}. 
In Ref.~\citen{hashimoto}, it was found that the dominant charge correlation function is interchange from the h-stripe CO type to the 3-fold CO type by the photo-irradiation,
while the dominant correlation for the v-stripe case is only weakened. 
%
The dominant difference from the previous calculations is the existence of the intermediate time domain in the h-stripe case. 
This can be attributed to two facts: 1) the coupling to the lattice degree of freedom induces a time lag in the transient excited states,
and 2) the electron-lattice interaction generates an energy barrier between the photo-excited Franck-Condon state and the 3-fold CO state. 

A number of theoretical calculations for the photo-induced electron and lattice dynamics in the CO systems have been performed. 
Not only the photo-induced transitions from one CO to another CO but also transitions from CO to metal
were shown by analyzing the extended Hubbard-type Hamiltonians with the lattice degrees of freedom, modeling the actual molecular organic salts~\cite{onda,kawakami,yonemitsu2,gomi2}. 
In the present study, we adopt a more basic model, i.e., the spinless fermion model coupled to a simplistic lattice degree of freedom via just the Holstein coupling,
and focus, in particular, on combination effects of the charge frustration and the electron-lattice coupling on the photo-induced transient dynamics. 
One main finding in the present study is the existence of the intermediate time domain, appearing in the h-stripe case. 
It is worth noting that the electron and lattice sectors show cooperative time evolutions in this time domain 
despite the fact that the original energy scales in the two sectors are much different with each other, i.e. $\omega_{lat}< t_{ij}, V_{ij}$. 
This is attributable to the frustration effect in the photo-excited state, which introduces a small energy scale in the electron sector. 
This interpretation is justified by the results that the electron and lattice sectors do not evolve cooperatively in the v-stripe case where the frustration effect is weak.  
%
%

We thank M.~Naka, J.~Nasu, K.~Iwano, and S.~Iwai for their helpful discussions.
This work was supported by JST CREST, 
JSPS KAKENHI Grant Numbers 26400377 and 26287070,
and the RIKEN iTHES project.
Some of the numerical calculations were performed using the supercomputing facilities at ISSP, the University of Tokyo.


\begin{thebibliography}{9}

\bibitem{yamamoto}
K. Yamamoto, S. Iwai, S. Boyko, A. Kashiwazaki, F. Hiramatsu, C. Okabe, N. Nishi, and K. Yakushi
J. Phys. Soc. Jpn. {\bf 77}, 074709 (2008).

\bibitem{tomioka}
Y. Tomioka and Y. Tokura,
Phys. Rev. B {\bf 70}, 014432 (2004). 


\bibitem{axe}
J. D. Axe, A. H. Moudden, D. Hohlwein, D. E. Cox, K. M. Mohanty, A. R. Moodenbaugh, and Y. Xu,
Phys. Rev. Lett. {\bf 62}, 2751 (1989). 

\bibitem{anderson}
P. W. Anderson,
Phys. Rev. {\bf 102}, 1008 (1956).

\bibitem{verwey}
E. J. W. Verwey, and P. W. Haaymann, 
Physica {\bf 8}, 979 (1941).

\bibitem{kawakami}
Y. Kawakami, T. Fukatsu, Y. Sakurai, H. Unno, H. Itoh, S. Iwai, T. Sasaki, K. Yamamoto, K. Yakushi, and K. Yonemitsu,
Phys. Rev. Lett. {\bf 105}, 246402 (2010). 

\bibitem{ishikawa}
T. Ishikawa, Y. Sagae, Y. Naitoh, Y. Kawakami, H. Itoh, K. Yamamoto, K. Yakushi, H. Kishida, T. Sasaki, S. Ishihara, Y. Tanaka, K. Yonemitsu, and S. Iwai,
Nat. Comm. {\bf 5}, 5528 (2014). 

\bibitem{onda}
K. Onda, S. Ogihara, K. Yonemitsu, N. Maeshima, T. Ishikawa, Y. Okimoto, X. Shao, Y. Nakano, H. Yamochi, G. Saito, and S. Koshihara,
Phys. Rev. Lett. {\bf 101}, 067403 (2008). 

\bibitem{gao}
M. Gao, C. Lu, H. Jean-Ruel, L. C. Liu, A. Marx, K. Onda, S. Koshihara, Y. Nakano, X. F. Shao, T. Hiramatsu, G. Saito, H. Yamochi, R. R. Cooney, G. Moriena, G. Sciaini, and R. J. D. Miller,
Nature, {\bf 496}, 343 (2013). 

\bibitem{beaud}
P. Beaud, A. Caviezel, S. O. Mariager, L. Retting, G. Ingold, C. Dornes, S-W. Huang, J. A. Johnson, M. Radovic, T. Huber, 
T. Kubacka, A. Ferrer, H. T. Lemke, M. Chollet, D. Zhu, J. M. Glownia, M. Sikorski, A. Robert, H. Wadati, M. Nakamura, M. Kawasaki, Y. Tokura, S. L. Johnson, and U. Staub,
Nat. Mater. {\bf 13}, 923 (2014). 

\bibitem{yonemitsu}
K. Yonemitsu and N. Maeshima, 
Phys. Rev. B {\bf 76}, 075105 (2007). 

\bibitem{dagotto}
J. Rinc\'{o}n, L. A. Al-Hassanieh, A. E. Feiguin, and E. Dagotto,
Phys. Rev. B {\bf 90}, 155112 (2014).

\bibitem{yonemitsu2}
K. Yonemitsu, 
Crystals, {\bf 2}, 56 (2012).

\bibitem{merino} 
J. Merino, H. Seo and M. Ogata, Phys. Rev. B {\bf 71}, 125111 (2005).

\bibitem{hotta}
C. Hotta, N. Furukawa, A. Nakagawa, and K. Kubo,
J. Phys. Soc. Jpn. {\bf  75}, 123704 (2006). 

\bibitem{nishimoto0}
S. Nishimoto, M. Shingai, and Y. Ohta, 
Phys. Rev. B, {\bf 78}, 035113 (2008). 

\bibitem{nishimoto}
S. Nishimoto, and C. Hotta,
Phys. Rev. B {\bf 79}, 195124 (2009).

\bibitem{naka}
M. Naka, and H. Seo,
J. Phys. Soc. Jpn. {\bf 83}, 053706 (2014).

\bibitem{hashimoto}
H. Hashimoto, H. Matsueda, H. Seo, and S. Ishihara, 
J. Phys. Soc. Jpn. {\bf 83}, 123703 (2014). 

\bibitem{park}
T. J. Park, and J. C. Light,
J. Chem. Phys. {\bf 85}, 5870 (1986).

\bibitem{prelovsek}
P. Prelovsek, and J. Bonca,
arXiv:1111.5931.

\bibitem{gomi2}
H. Gomi, T. Kawatani, T. J. Inagaki, and A. Takahashi, 
J. Phys. Soc. Jpn. {\bf 83}, 094714, (2014). 

\bibitem{initial}
In the initial states for the Newtonian equation, small random lattice distortions ($\delta q_{i}$) and momenta ($\delta p_{i}$) are introduced in order to break a high symmetry in a cluster, in which a condition $\omega_{lat}(\delta q_{i}^{2}+\delta p_{i}^{2})/2 = 10^{-6} t$ is satisfied at each site. 



%
%
%
%
%




%
%
%
%
%



    










\end{thebibliography}
\end{document}